\def\cm{cm$^{-1}$}
\def\be{\begin{equation}}
\def\ee{\end{equation}}

\documentclass[twocolumn,prb,showpacs,amsmath,amssymb]{revtex4}

\usepackage{graphicx}

\usepackage{dcolumn}

\usepackage{bm}
\begin{document}

\title{Signatures of electron-electron correlations in the optical spectra
of $\alpha$-(BEDT-TTF)$_2M$Hg(SCN)$_4$ ($M$=NH$_4$ and Rb)}

\author{N. Drichko$^{1}$}
\author{M. Dressel$^{1}$}

\author{A. Greco$^2$}
\author{J. Merino$^3$}
\author{J. Schlueter$^4$}
\affiliation{
$^1$ 1.~Physikalisches Institut, Universit{\"a}t Stuttgart, Pfaffenwaldring 57, 70550 Stuttgart, Germany \\
$^2$ Dept. de Fisica Teorica de la Materia Condensada, Universidad
Autonoma de Madrid, Spain\\
$^3$ Facultad de Ciencias Exactas Ingenier\'ia y Agrimensura e Instituto de F\'isica Rosario (UNR-CONICET), Rosario, Argentina\\
$^4$ Material Science Division, Argonne National Laboratory,
Argonne, U.S.A}

\begin{abstract}
We interpret the optical spectra of
$\alpha$-(BEDT-TTF)$_2M$Hg(SCN)$_4$ (M=NH$_4$ and Rb) in terms of
a 1/4 filled metallic system close to charge ordering and show
that in the conductivity spectra of these compounds a fraction of
the spectral weight is shifted from the Drude-peak to higher
frequencies due to strong electronic correlations. Analyzing the
temperature dependence of the electronic parameters, we
distinguish between different aspects of the influence of
electronic correlations on optical properties. We conclude, that
the correlation effects are slightly weaker in the NH$_4$ compound
compared to the Rb one.

PACS numbers:
74.70.Kn    
72.15.Nj,  
\end{abstract}

\maketitle

\section{INTRODUCTION}
The physical properties of layered molecular conductors are known
to be influenced by electron-electron interactions. The
comparative strength of the electronic correlations can be tuned
by small modifications of the chemical composition of the
crystals, that might result in different ground states for
isostructural compounds. In addition, the behavior of the
correlated systems depends essentially on the conduction-band
filling \cite{McKenzie97,Drichko05}; while for half-filled systems
the on-site correlations $U$ are of importance, the quarter-filled
systems stay metallic at any values of $U$, if the inter-site
correlations $V=0$.

The importance of electronic correlations in 1/4-filled
two-dimensional organic conductors was investigated for the
$\theta$-\-(BEDT-TTF)$_2$\-$MM^{\prime}$\-(SCN)$_4$
family\cite{Mori98,Mori98a}, where a charge-order metal-insulator
transition is observed at the temperature decrease. On the other
hand, extensive theoretical studies were performed for the
electronic correlations in 1/4-filled layered molecular metals
close to the charge-order
transition\cite{Merino01,Calandra,Merino03,Dressel03}. Here we
present an optical study of 1/4 filled metals
$\alpha$-(BEDT-TTF)$_2M$Hg(SCN)$_4$ ($M$=NH$_4$,Rb), aimed on
obtaining the effective mass and scattering rate of charge
carriers, in order to compare them with theory and to distinguish
between the electronic correlations and other factors defining
their behavior. While in this family of compounds is known that
superconductivity ($T_c=1$~K) competes with a density-wave
($T_c=8$-10~K) ground state, we argue that the properties at
temperatures between 300 and 6~K are determined by the distinct
ratios of inter-site Coloumb repulsion $V$ to the bandwidth.

\section{RESULTS}
The polarized reflectivity of single crystals of
$\alpha$-(BEDT-TTF)$_2M$Hg(SCN)$_4$ ($M$=NH$_4$, Rb) was measured
in the conducting plane along the main optical axes (parallel and
perpendicular to the stacks of BEDT-TTF molecules) in the
frequency range between 50 and 6000~\cm. The sample was cooled
down to 6~K with a rate of 1~K/min; spectra were taken at 300,
200, 150, 100, 50 and 6~K. In order to measure the absolute values
of reflectivity, the sample was covered {\it in situ} with 100~nm
and used as a reference; this technique is described in Ref.~[9].

In agreement with the calculated higher values of transfer
integrals between the stacks of BEDT-TTF
molecules\cite{Mori_Term}, the reflectivity is always higher in
the direction of the $E\bot$stacks. The reflectivity increases on
cooling for both salts, as expected for metals. In the
conductivity spectra, the wide finite-frequency electronic
features are observed in addition to a comparatively narrow
zero-frequency peak. Furthermore, the narrow features in the
400-1600~\cm\ range near the frequencies of the BEDT-TTF
intramolecular vibrations are known to be the $a_g$ vibrations of
BEDT-TTF activated by the electron-molecular vibrational
coupling\cite{review}; they are out of the scope of this paper.

\begin{figure}[h]
\begin{center}
\includegraphics[width=11.0cm]{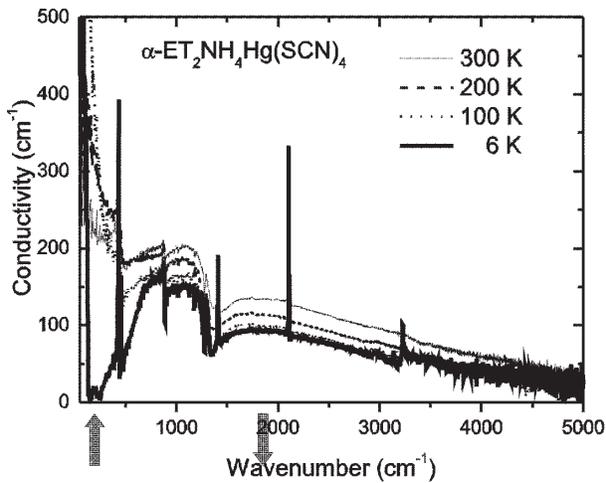}
 \caption{Optical conductivity of $\alpha$-(BEDT-TTF)$_2$NH$_4$Hg(SCN)$_4$ in E$\parallel$stacks polarization.
 The arrows indicate the increase of Drude spectral weight and the decrease of high-frequency features on
 cooling.} \label{fig:NH4}
\end{center}

\end{figure}

\section{DISCUSSION}
To follow the electronic spectral features we performed a
Drude-Lorentz fit of the experimental data. In both polarizations
the spectra of these compounds are described well by a Drude peak,
a maximum at about 1000~\cm\, and an overdamped maximum at about
2400~\cm. The high-frequency maximum can be distinguished in the
whole temperature range; the frequency does not change with
temperature. The low-frequency maximum is separated  from the
Dude-peak by a pseudogap only below $T=200$~K when the Drude-peak
becomes narrow enough, though the fit suggests that the maximum
exists already at room temperature. On cooling down to 6~K this
feature shifts from 1000 to 800~\cm.

\begin{figure}[h]
\begin{center}
\includegraphics[width=11.0cm]{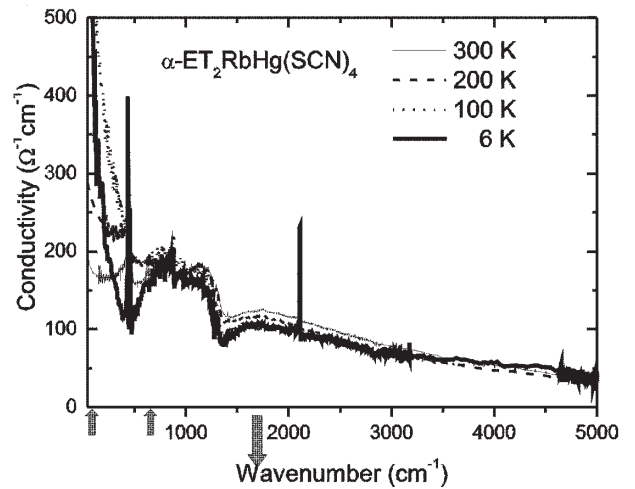}
 \caption{Optical conductivity of $\alpha$-(BEDT-TTF)$_2$RbHg(SCN)$_4$ in E$\parallel$stacks polarization.
 The arrows indicate a weak increase of Drude spectral weight and the
 1000~\cm\ feature,
 and the decrease of the mid-infrared feature on cooling.}\label{fig:Rb}
 \end{center}
\end{figure}
The observed bands cannot be interpreted as interband transitions,
according to the band structure calculations\cite{Mori_Term},
since they do not occur at known transition energies. We attribute
the observed electronic features to the spectral weight shifted
from the Drude-peak to higher frequencies due to the influence of
electronic correlations; a similar interpretation was proposed for
the spectra of $\theta$-phase compounds\cite{Merino03,Tajima00},
where a very intense feature and a pseudogap at 300~\cm\ was
previously observed in the low temperature spectra of
$\alpha$-(BEDT-TTF)$_2K$Hg(SCN)$_4$ and interpreted as an
indication of the fluctuating charge ordering\cite{Dressel03}.

In a one-band approximation, these three electronic features
together give us the total spectral weight\cite{DresselGruner02}
 $
\omega^2_p=8\int_0^{\omega_c}\sigma_1(\omega)\,{\rm d}\omega
\label{sw} $ received by the integration up to frequency
$\omega_c$. The minimum in the conductivity spectrum indicates the
exhaustion of the band, i.e.\ the band width  $W$. Consequently,
the contraction of the crystal on cooling, which changes the size
of the transfer integrals, results in a change of the total
spectral weight, as estimated by the one-dimensional tight-binding
approximation:
\begin{eqnarray}
\omega_p^2&=&\frac{16td^2e^2}{\hbar^2
V_m}\sin\left\{\frac{\pi}{2}\rho\right\} \label{eq:plasmaparallel}
\quad .
\end{eqnarray}

The calculated temperature dependence of the transfer
integrals\cite{Ono97} shows a striking in-plane anisotropy: about
a 7\%\ increase of the inter-stack transfer integrals is expected
on cooling the crystals from 300 down to 4~K, while  two of the
three in-stack transfer integrals stay constant. In agreement with
it, the spectral weight estimated from our optical experiments
does not change on cooling for the polarization
$E$$\parallel$stacks, but a considerable increase is observed in
the $E\bot$stacks direction: for the Rb-compound  it is about
10\%, in agreement with the $t$ change, while the increase of the
spectral weight for the NH$_4$ compound amounts up to 30\%, well
above the expectations. In addition, the increase of the transfer
integrals leads to a decrease of the $V/t$ ratio that defines the
distribution of the spectral weight between the Drude-like peak
and high frequenies\cite{Merino03}. Indeed, a pronounced shift of
the spectral weight from the mid-infrared to the zero-frequency
peak is observed in the spectra of polarization $E\bot$stacks.

Neither the total spectral weight, nor the hopping parameters
change in the E$\parallel$stacks direction. Therefore, the
observed shift of the spectral weight from the high-frequency
features to the Drude peak on cooling (indicated by arrows in Fig.
1, 2) is consistent with predictions of the charge fluctuation
scenario\cite{Merino03}. A discussion of the main ingredients of
this theory is revised in the following section.

\section{Comparison to charge fluctuation scenario}

The simplest model to describe the relevant electronic properties
of the $\alpha$-type salts is a one band extended Hubbard model at
one-quarter filling on the square lattice. This model contains a
kinetic energy scale given by the hopping, $t$, the on-site
Coulomb repulsion, $U$ and the intersite Coulomb repulsion, $V$.
This last Coulomb repulsion accounts for the Coulomb repulsion
between two electrons on the two nearest-neighbors molecules. From
quantum chemistry it is known that $U$ is larger than the
bandwidth, $W=8t$ whereas $V$ is comparable to the hopping and
therefore cannot be neglected. In fact the model contains a
transition from a metal to a charge ordered state induced by $V$,
when $V=2t$ (Ref.~\cite{Calandra}) at zero temperature. Close to
the charge ordering transition it is found from exact
diagionalization calculations that: (i) the Drude weight is
rapidly suppressed; (ii) the optical spectral weight is
redistributed so that a net shift from low frequencies into a
mid-infrared band occurs; (iii) a characteristic low frequency
feature appears in the optical spectra as a consequence of the
presence of the strong charge fluctuations occurring in the
metallic phase close to the transition.

At non-zero temperatures, exact diagionalization cannot be used
reliably due to the system size, and so one needs to rely on other
theoretical methods. Slave-boson, CPA (Coherent Potential
Approximations) and large-$N$ calculations on the extended Hubbard
model predict that sufficiently close to $V_c$ and within the
metallic phase, increasing the temperature drives the system into
the charge ordered phase. This reflects the unusual 're-entrant'
character of the metal-charge ordered transition, which leads to
an interesting temperature dependence of the electronic properties
of the metal when being close to a checkerboard charge ordering
transition at T=0 such as an increase in the electron effective
mass with increasing temperature \cite{Merino03}. This behavior is
different from the one found in a conventional metal in which
increasing the temperature would tend to suppress the
renormalization of the electrons due to their interaction with
other electrons, phonons, etc, effectively decreasing their
effective mass enhancement as $T \rightarrow T_F$, where $T_F$ is
the Fermi temperature of the metal. Apart from the increasing
effective mass with increasing temperature characteristic of the
charge ordering transition, the scattering rate, $1/\tau(T)
\propto T^2$ behavior for $T<T^*$  becomes linear: $1/\tau(T)
\propto T$ for $T>T^*$ due to the onset of the charge fluctuations
just like in the case of the interaction of electrons with other
boson-like excitation such as phonons\cite{Ashcroft}. The
temperature $T^*$ defines the distance of the metal from the
critical point at which the charge ordering transition occurs.
Hence, $T^* \rightarrow 0$ as $V \rightarrow V_c$. In a few words,
$T^*$ is the characteristic energy scale describing the onset of
the charge fluctuations. This energy scale is then very small:
$T^*<<T_F$ whenever the system is sufficiently close to the charge
ordering transition.

Summarizing, the charge ordering scenario predicts for a metal
close to the charge ordered phase ($V \lesssim V_c$): (i) a
suppression of the coherent part of quasiparticles compared to the
electrons of the non-interacting ($V=0$) system, (ii) the
destruction of the quasiparticles becoming more incoherent as the
temperature is gradually increased above $T>T^*$ from $T=0$, (iii)
a linear dependence of the scattering rate: $1/\tau(T)$ for
$T>T^*$. The slope of $1/\tau(T)$ increases as the system is
driven closer to the transition.

\begin{figure}
\begin{center}
\vspace*{-5mm}
\includegraphics[width=10.0cm]{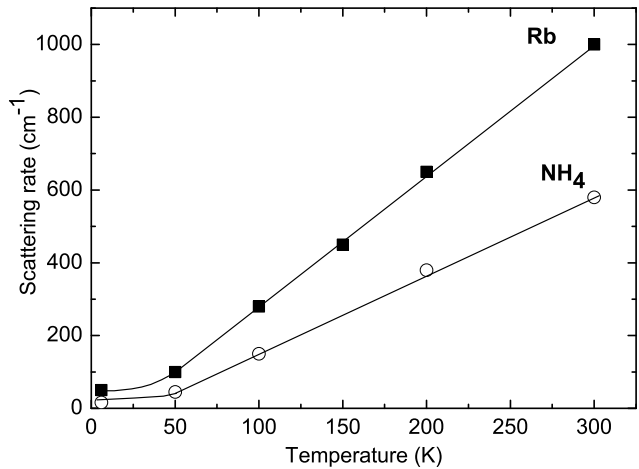}
 \caption{Temperature dependence of Drude scattering rate in
 E$\parallel$stack direction for the NH$_4$ and Rb salt. $T^*$ is calculated assuming $t$=0.060
eV.}\label{fig:GammaD}
\end{center}
\end{figure}

Indeed, this explains the experimentally observed  decrease of the
Drude weight with increasing temperature in the spectra of
E$\parallel$stacks direction in these compounds. The effective
mass of the charge carriers $m_D$ was calculated from the Drude
plasma frequency $\omega_p$: $
 \omega_{p} =
\left(\frac{4\pi Ne^2}{m_D}\right)^{1/2} \quad \label{eq:defplasma2}$.
The observed decrease of the plasma frequency with temperature can
either be caused by an enhancement of the effective mass or a decrease
of the number of carriers. Since the later one seems unlikely, we
assume the carrier concentration to be constant with temperature and
interpret the observed change of $\omega_p$ as an increase in the
effective mass. For the Rb compound $m^*$ increases from $4.2m_e$ to
$5.1m_e$ and from $4.1$ to $4.9$ for the NH$_4$ compound with
increasing the temperature from 6 to 300 K is observed, in agreement
with prediction (ii) of the charge fluctuating scenario. Also, the
temperature behavior of the Drude scattering rate
(Fig.~\ref{fig:GammaD}), though characterized by only six points, is
distinctly  linear with temperature down to about 50 K, while below
this temperature it changes to a much slower dependence in agreement
with prediction (iii).

The absolute yields of the scattering rate of the Drude-component
are higher in the Rb compound, resulting in the presence of some
spectral weight in the pseudogap at about 300 \cm. In addition, it
shows a much larger slope in the scattering rate, therefore we
conclude from point (iii) that the Rb compound closer is to the
charge ordering transition than the NH$_4$ salt. Other
explanations for the different scattering rates would be difficult
since the compounds are isostructural and have very similar
properties. For instance, a different number of defects in the two
salts would give a constant shift in the values of scattering rate
between the two compounds but would have difficulties in
explaining the different observed slopes. Phonons could also give
changes in the temperature dependence however at much larger
temperatures than the charge fluctuation scale, $T^*$ as the Debye
frequency is expected to be of the order $T_F$. The magnitude of
the characteristic temperature scale for the charge fluctuations,
$T^*$ is theoretically of about $T^* \approx 0.2 t$. Taking
estimates for $t=0.06$ eV, the estimates for $T^*$ are of the
right order of magnitude compared to the experiments.

\section{CONCLUSIONS}
The optical investigations of $\alpha$-BEDT-TTF$_2M$Hg(SCN)$_4$
(M=NH$_4$ Rb) compounds reveal the anisotropic behavior of the
crystals upon lowering the temperature. Perpendicular to the
stacks, the total spectral weight increases on cooling with the
increase of the overlap integrals $t$, while the respective
decrease of the $V/t$ ratio leads to the shift of the spectral
weight from the finite-frequency features to the Drude-peak. No
increase of the total spectral weight is observed parallel to the
stacks, as these transfer integrals do not change on cooling. We
attribute the weak shift of the spectral weight and an increase of
the effective mass with temperature to the 're-entrant' character
of the charge ordering transition driven by the Coulomb repulsion
$V$ by which the critical $V_c$ is decreased with increasing
temperature. The larger slope of the scattering rate of the charge
carriers and the larger effective masses of the Rb compound
compared to NH$_4$, suggests that the Rb salt is closer to the
charge ordering transition than the NH$_4$ salt. This conclusion
is based on a comparison with the predictions of the properties of
metals under the effect of charge fluctuations.

\section*{ACKNOWLEDGMENTS}

The work was supported by the Deutsche For\-schungs\-gemeinschaft
(DFG). N.D. is grateful for the support of Alexander von Humboldt
Foundation.


\begin{thebibliography}{9}
\bibitem{McKenzie97} R.H. McKenzie,  {\it Science} {\bf 278}, 821 (1997).
\bibitem{Drichko05}N. Drichko, K. Petuhov, M. Dressel, O. Bogdanova, E. Zhilyaeva, R. Lyubovskaya, A. Greco, J. Merino, {\it Phys.Rev.B} {\bf 72}, 024524
(2005).
\bibitem{Mori98} H. Mori, S. Tanaka, and  T. Mori,  {\it Phys. Rev. B} {\bf 57}, 12023 (1998).
\bibitem{Mori98a}  T. Mori {\it Bull. Chem. Soc. Jpn.} {\bf 71}, 2509 (1998).
\bibitem{Merino01} J. Merino and R. H. McKenzie. {\it Phys. Rev. Lett.} {\bf 87}, 237002 (2001) .
\bibitem{Calandra} M. Calandra, J. Merino and R. H. McKenzie, {\it Phys. Rev. B} {\bf 66}, 95102 (2002).
\bibitem{Merino03}J. Merino, A. Greco, R. McKenzie, M.Calandra,
{\it Phys. Rev. B} {\bf 68}, 245121 (2003).
\bibitem{Dressel03}M. Dressel, N. Drichko, J. Schlueter, and J. Merino.
{\it Phys.Rev.Lett.} {\bf 90}, 167002 (2003).
\bibitem{Homes93}C.C.  Homes,  M. Reedyk, D.A. Cradles, and T. Timusk,
{\it Appl. Opt.} {\bf 32}, 2976 (1993).
\bibitem{Mori_Term}T. Mori, H. Inokuchi, H. Mori, S. Tanaka, M. Oshima, G. Saito
{\it J. Phys. Soc. Jpn} {\bf 59}, 2624 (1990) .
\bibitem{review}M. Dressel and N. Drichko.
{\it Chem. Rev.} {\bf 104}, 5689 (2004).
\bibitem{Tajima00}Tajima,
H.; Kyoden, S.; Mori, H. and  Tanaka, S. {\it Phys. Rev. B} {\bf
62}, 9378 (2000).
\bibitem{DresselGruner02}M.~Dressel and G.~Gr\"{u}ner, {\it Electrodynamics of Solids}  (Cambridge University Press, Cambridge, 2002).
\bibitem{Ono97}S. Ono, T. Mori, S. Endo, N. Toyota, T. Sasaki, Y. Watanabe, T. Fukase
{\it Physica~C} {\bf 290}, 49 (1997).
\bibitem{Ashcroft} N. W. Ashcroft {\it Solid State Physics} (Saunders
College, Philadelphia, 1976)
\end{thebibliography}
\end{document}